\documentclass[useAMS,usenatbib,usegraphicx]{mn2e}

\usepackage{graphicx}
\usepackage[latin1]{inputenc}
\usepackage{color}
\usepackage{times}
\usepackage{natbib}
\usepackage{setspace}
\newif\ifAMStwofonts
\AMStwofontstrue
\definecolor{red}{rgb}{1,0.,0.}

\newcommand{\morgana}{{\sc morgana}}
\newcommand{\munich}{WDL08}

\def\lesssim{\lower.5ex\hbox{$\; \buildrel < \over \sim \;$}}
\def\gtrsim{\lower.5ex\hbox{$\; \buildrel > \over \sim \;$}}

\title[Bulgeless Galaxies] {The other side of Bulge Formation in a
  $\Lambda$CDM cosmology: \\ Bulgeless Galaxies in the Local Universe}

\author[Fontanot et al.]{
  \parbox[t]{\textwidth}{ Fabio Fontanot$^1$, Gabriella De Lucia$^1$,
    David Wilman$^2$ and Pierluigi Monaco$^{3,1}$}
    \vspace*{6pt}\\
    $^1$ INAF-Osservatorio Astronomico di Trieste, Via Tiepolo 11, I-34131 Trieste, Italy \\
    $^2$ MPE Max-Planck-Institute f\"ur Extraterrestrische Physik, Giessenbachstra\ss e, D-85748 Garching-bei-M\"unchen, Germany \\
    $^3$ Dipartimento di Fisica, sez. Astronomia, via G.B. Tiepolo 11, I-34131 trieste, Italy \\
    email: fontanot@oats.inaf.it}

\begin{document}
\date{Accepted ... Received ...}

\maketitle

\begin{abstract} 
We study the physical properties, formation histories, and environment
of galaxies without a significant ``classical'' spheroidal component,
as predicted by semi-analytical models of galaxy formation and
evolution. This work is complementary to the analysis presented in
\citet{DeLucia11}, where we focus on the relative contribution of
various physical mechanisms responsible for bulge assembly in a
$\Lambda$CDM cosmology. We find that the fraction of bulgeless
galaxies is a strong decreasing function of stellar mass: they
represent a negligible fraction of the galaxy population with
$M_\star>10^{12} M_\odot$, but dominate at $M_\star<10^{10}
M_\odot$. We find a clear dichotomy in this galaxy population, between
central galaxies of low-mass dark matter haloes, and satellite
galaxies in massive groups/clusters. We show that bulgeless galaxies
are relatively young systems, that assemble most of their mass at
low-redshift, but they can also host very old stellar
populations. Since galaxy-galaxy mergers are assumed to lead to the
formation of a spheroidal component, in our models these galaxies form
preferentially in low-mass haloes that host a small number of
satellites galaxies. We show that the adopted modelling for galaxy
mergers represents a key ingredient in determining the actual number
of bulgeless galaxies. Our results show that these galaxies are not a
rare population in theoretical models: at $z\sim0$, galaxies with no
classical bulge (but often including galaxies with the equivalent of
pseudo-bulges) account for up to $14\%$ of the galaxies with $10^{11}
< M_\star/M_\odot < 10^{12}$.
\end{abstract}


\section{Introduction}\label{sec:intro}
Since the introduction of the morphological classification scheme by
\citet{Hubble1926}, two components are traditionally identified in
galaxies: a centrally concentrated spheroidal-like structure
(``bulge'') and a disc-like stellar distribution, often associated
with spiral arms. The galaxy population can be (and usually is)
classified according to the relative contribution of these two
components to the total light emitted by the system. A finer
classification takes into account the contribution from other
components (e.g. bars, spiral arms). In the last few decades,
observational evidence has been gathered to indicate that this picture
is oversimplified: bulges (which contribute up to $60\%$ of the
stellar mass in massive galaxies in the local Universe,
\citealt{Gadotti09}) are now seen as a heterogeneous class, including
purely spheroidal systems (elliptical galaxies), ``classical'' bulges
(dynamically and photometrically similar to ellipticals, but with
significant kinematical differences, see
e.g. \citealt{DaviesIllingworth83}), and ``pseudo'' bulges
(characterised by ``disc''-like exponential profiles or kinematics,
see e.g. \citealt{KormendyKennicutt04} and references herein).

Theoretical models predict that early star formation takes place
mainly in discs that form due to the conservation of the angular
momentum acquired through early torques acting during the
proto-galactic stage. Bulges form as the result of physical processes
able to remove angular momentum from stars and gas. In particular,
classical bulges and purely spheroidal systems are believed to be
associated with the most violent dissipative processes, like mergers
and close interactions. On the other hand, pseudo bulges are usually
linked to secular evolution of gravitational instabilities in the disc
component: the ``unstable'' structure is expected to find a new
equilibrium following a rearrangement of part of the gas and stars in
a central structure with enhanced density. Whenever these processes
are infrequent and/or inefficient, we expect the galaxy morphology to
be dominated by a disc component. These galaxies are often referred to
as ``bulgeless'', and have been seen as a potential challenge for
current theories of galaxy formation and evolution in a $\Lambda$CDM
Universe (see e.g. \citealt{DOnghiaBurkert04}). Following these
suggestions several authors (see e.g. \citealt{GrahamWorley08}) used
bulge-disc decomposition algorithms to determine the relative
contribution of these two components in samples of local galaxies. In
particular, \citet{Weinzirl09} considered a sample of 143 bright
($M_B<-19.3$) low inclination spirals and found that a relevant
fraction ($\sim66\%$) of these is dominated by a disc component
accounting for more than $80\%$ of their total stellar mass. Starting
from a sample of $\sim 4000$ bright ($M_B<-18$) galaxies at
$0.013<z<0.18$, \citet{Cameron09} computed a total stellar mass
density for pure-disc objects of $1.3 \pm 0.1 M_\odot Mpc^{-3}$. More
recently, \citet[K10 hereafter]{Kormendy10} consider a sample of 19
relatively massive (rotation velocities $V_{\rm rot}>150 \, {\rm
  km/s}$) and close ($<8$ Mpc) galaxies, and find that 4 of these are
consistent with being pure disc galaxies. 7 galaxies in the same
sample (including the Milky Way) have pseudo-bulges. In total, K10
estimate that $58-74\%$ of the galaxies in their sample do not
experience relevant mergers in the past (they include in the bulgeless
category those galaxies with a pseudo-bulge). Considering that, at
these masses, quiet merger histories are rare, all these authors
propose their estimates as a crucial test for galaxy formation models.

The formation of bulgeless galaxies represents a classical challenge
also for cosmological N-body hydrodynamical simulations. If conserved
during its collapse toward the centre of the galaxy, the angular
momentum of the gas is sufficient to produce large discs (e.g.,
\citealt{FallEfstathiou80, MoMaoWhite98}). Cooling in small and dense
progenitors at high redshift, however, condense the gas in their inner
regions. Dynamical friction on the orbiting satellites then dissipates
the gas angular momentum (e.g., \citealt{DOnghia06}). As a result,
discs in N-body simulations are too compact (with rotation curves that
usually peak at a few kpc) with respect to observational measurements
(\citealt{SteinmetzNavarro99}; see \citealt{Mayer08} for a
review). One of the proposed (and so far most successful) solution to
the angular momentum `catastrophe' requires a fine tuning of the
feedback in small high-redshift progenitors so as to avoid early gas
cooling (e.g., \citealt{Governato10}). It should be noted that loss of
angular momentum is caused also by secular evolution of discs and bar
instability (see e.g. \citealt{Debattista06}), and that these are more
easily triggered in discs that are made compact by partial loss of
angular momentum \citep{Curir06}.

In a recent paper, \citet[hereafter Paper I]{DeLucia11} analyse the
predictions from semi-analytical models (SAMs) of galaxy formation and
evolution within a $\Lambda$CDM cosmology, and quantify the relative
contribution of different processes (major and minor mergers, and disc
instabilities) to the assembly of bulges. In this paper, we will
tackle a complementary question, i.e. determine under which conditions
a model galaxy does {\it not} develop a significant spheroidal
component.

\section{Models}\label{sec:models}

In this paper, we present predictions from two independently developed
SAMs, namely the \citet[hereafter \munich]{Wang08} implementation of
the ``Munich'' model and the \morgana~model, as adapted to the WMAP3
cosmology in \citet{LoFaro09}. We refer to the original papers for a
detailed discussion of the models used in this study, and to Paper I
for a detailed description of the recipes adopted to model bulge
formation. In this section, we provide a brief summary of these
ingredients.

Both models consider similar channels leading to the assembly of a
spheroidal component, namely galaxy-galaxy mergers and disc
instabilities. There are however, significant differences in the
treatment of these processes. In both models, {\it major mergers}
completely destroy the disc components of the two galaxies. The
remnant spheroidal galaxy may eventually regrow a disc, if fed by an
appreciable cooling flow at later times. During {\it minor mergers},
both SAMs assign the stellar component of the secondary galaxy to the
bulge of the remnant, but make different assumptions about the stars
formed during the burst associated with the merger: \munich~gives them
to the disc of the remnant galaxy, while \morgana~gives them to its
bulge. This different choice implies that the contribution of minor
mergers to bulge assembly is more important in
\morgana. \citet{DeLucia10} compared the merger models implemented in
\morgana~and \munich~and found that the former provides {\it merger
  times} that are systematically shorter than those used in the latter
(by about an order of magnitude): this translates into a higher
frequency of merger events in \morgana~and has important consequences
for the timing of bulge formation, as discussed in Paper I.  In
particular, due to the shorter merger times and the different
treatment of minor mergers, we have shown that \morgana~predicts a
larger stellar mass locked in bulges at each redshift, and larger mean
bulge-to-total ratios for galaxies of all masses.

The two models used in this study also differ in their treatment of
{\it disc instabilities}: both models adopt the stability criterion
defined in \citet{Efstathiou82}, but use different definitions for the
relevant physical quantities (in particular disc velocities and radii,
see sec. 7 and Fig.~10 of Paper I). In addition, they make different
assumptions about the re-arrangement of baryons following instability
events: in \munich~only the stellar mass fraction necessary to restore
stability is transferred from the disk to the bulge. In \morgana, a
significant fraction (i.e. half) of the baryonic mass (both gas and
stars) of the disk is transferred to the bulge. As shown in Paper I,
the approach adopted in the \morgana~model translates into a more
prominent role of the disk instability channel in bulge formation.

Following Paper I, we consider in the following three different
implementations for each model: a {\it standard} implementation, which
includes both mergers and disc instability, a {\it pure merger}
implementation where we switch off the disk instability channel, and a
model that adopts the \citet[{\it HOP09} hereafter]{Hopkins09b}
prescriptions for the re-distribution of gas and stars during mergers,
and for modelling the fraction of disc material which survives merger
events. Briefly, the HOP09 approach reduces the efficiency of bulge
formation and increases the fraction of baryonic mass in disc
components at each redshift. In addition, it assumes that a fraction
of the disc survives even during major mergers. In Paper I, we showed
that this change has important consequences for the predicted space
density of purely spheroidal (elliptical) galaxies, but it does not
affect bulge formation in galaxies less massive than $\sim 10^{10}
M_\odot$.

Our SAMs do not allow a fine classification of the different bulge
subclasses to be made. In particular, it is not possible to
disentangle between the formation of a pseudo or classical bulge,
based just on the properties of the final spheroidal component.
Nonetheless, our strategy provides a natural framework for the
analysis presented in this study: assuming that classical bulges are
associated with mergers, and that pseudo-bulges originate from
instabilities, the standard implementation gives the full statistics
for disc-dominated galaxies, while the comparison with the pure merger
implementation provides information about the relative contribution of
classical and pseudo-bulges.

In this paper we will define as ``bulgeless'' all model galaxies with
a bulge-to-total ($B/T$) mass ratio lower than $0.1$, i.e. all model
galaxies whose bulges contribute to less than $10\%$ of their total
stellar mass. In the following, we consider only galaxies with
$M_\star>10^9 M_\odot$, which is above the resolution limit of our
simulations. Resolution effects may play a role in our definition of
bulgeless galaxies, especially for galaxies with $M_\star<10^{10}
M_\odot$, since we are not able to account for mergers with and
between lower mass galaxies. We note, however, that galaxy formation
in lower masses haloes is an increasingly inefficient process (see
e.g. \citealt{BensonDevereux10}). So minor mergers with small galaxies
hosted in sub-resolution haloes are not expected to contribute much to
the growth of the central spheroid, and to affect the results
discussed below.

\section{Results}\label{sec:results}
\begin{figure}
  \centerline{ \includegraphics[width=9cm]{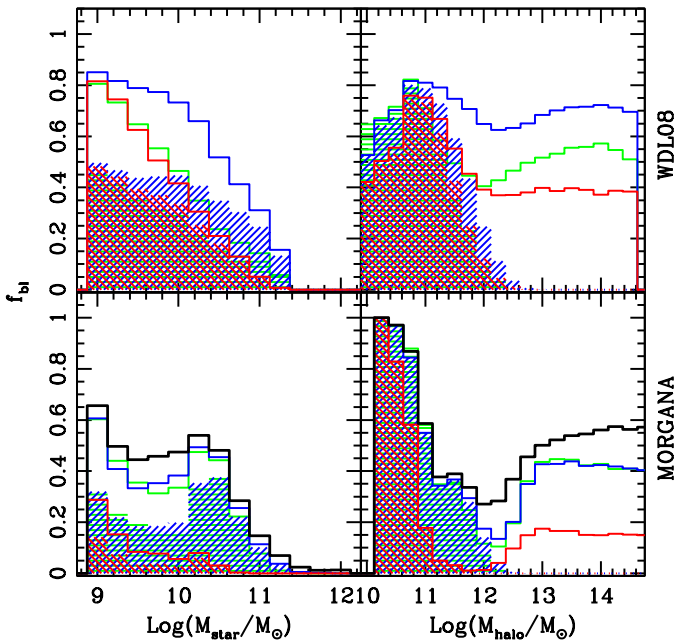} }
  \caption{The fraction of bulgeless galaxies ($f_{\rm bl}$) as a function of
    their stellar mass and parent halo mass. In each panel, red, blue and green
    lines refer to the standard model, to the pure merger implementation, and
    the HOP09 implementation respectively. Shaded histograms are for central
    galaxies, and use the same colour coding. In the panels showing results
    from \morgana, the thick solid black line shows results obtained adopting
    longer merger times (see text for more details).}\label{fig:mass}
\end{figure}

In this section, we will discuss in detail the properties, environment and
formation history of bulgeless galaxies, as predicted by our SAMs. In
particular, we consider the fraction $f_{\rm bl}(P)$ of model galaxies with
$B/T<0.1$ and with a given property $P$, and the normalised (to the total
number of systems) distribution $n_{\rm bl}(P)$ of bulgeless galaxies as a
function of $P$.

Fig.~\ref{fig:mass} shows $f_{\rm bl}(M)$ as a function of stellar
mass ($M_\star$) and parent halo mass ($M_{\rm DM}$). In each panel,
the red, blue and green histograms refer to the standard, pure merger
and HOP09 implementations respectively. The shaded histograms show the
contribution of central galaxies (same colour coding). Common trends
between the two SAMs and the three implementations considered can be
seen. In the standard runs, $f_{\rm bl}(M_\star)$ exhibits a marked
decrease as a function of $M_\star$ (left panels). A similar result is
obtained in the pure merger runs, although in this case the resulting
fractions are larger than those obtained in the standard models,
because of the lack of bulges forming via the disc instability
channel. Neglecting bulge formation via disc instabilities has a
stronger effect on predictions from \morgana~than on those from
\munich. This is due to the more efficient mass transfer associated
with this physical mechanism in the former model. At all masses,
$f_{\rm bl}(M_\star)$ in \munich~is larger than in \morgana~for both
implementations: this is due to the assumption of shorter merger times
in \morgana, that leads to a more efficient bulge formation in this
model with respect to \munich. Therefore, the relatively small $f_{\rm
  bl}(M_\star)$ in the standard \morgana~run is due to a combination
of shorter merger time scales, and stronger mass transfer via disc
instabilities. We test explicitly the influence of different merger
times, by re-running \morgana~using the same dynamical friction
prescription adopted in \munich~- results are shown as a black solid
line in the bottom panels of fig.~\ref{fig:mass}. As expected, when
longer merger times are assumed, $f_{\rm bl}(M)$ increases because
more galaxies are able to avoid mergers.

We find that the HOP09 implementations do not alter significantly the
fraction of bulgeless galaxies with respect to the runs considered
above: this is consistent with conclusions from our Paper I that this
recipe does not affect bulge formation in galaxies less massive than
$\sim 10^{10} M_\odot$. We note, however, that the HOP09 predictions
for the \munich~model are closer to the standard results, while for
\morgana~they are closer to the pure merger model results. This
difference is due to the different treatment of the stars associated
with bursts triggered by minor mergers: as explained above, these are
given to the disc of the remnant galaxy in the standard implementation
of the \munich~model, while when adopting the HOP09 recipes this model
assumes (as done in the \morgana~model) that these stars go to the
bulge of the remnant galaxy.

We then consider the fraction of bulgeless galaxies as a function of
their parent halo mass ($f_{\rm bl}(M_{\rm DM})$, right panels). A
clear dichotomy in the bulgeless population can be seen: lower-mass
haloes have an increasing probability of hosting a central bulgeless
galaxy (larger than $70\%$ for $M_{\rm DM}<10^{11} M_\odot$), while
almost all central galaxies of haloes with $M_{\rm DM} \gtrsim 10^{12}
M_\odot$ host significant bulges in all implementations. On the other
hand, bulgeless satellites constitute an important fraction of
cluster/group galaxies.  \morgana~predicts a lower $f_{\rm bl}(M_{\rm
  DM})$, and a clear dip at $M_{\rm DM} \sim 10^{12} M_\odot$
(depending on the chosen implementation). For the \munich~model the
trends as a function of the halo mass are somewhat weaker.

\begin{table}
\begin{center}
  \begin{tabular}{ccc}
    \hline
     & $10^{10}<\frac{M_\star}{M_\odot}<10^{11}$ & $10^{11}<\frac{M_\star}{M_\odot}<10^{12}$ \\
    \hline
    \multicolumn{3}{c}{Standard Implementation} \\
    \munich  & $26^{+13}_{-9} \, \%$ & $0^{+14}_{-0} \, \%$ \\
    K10 (total bulges) & $60^{+26}_{-33} \, \% $ & $27^{+25}_{-16} \, \% $ \\
    \hline
    \multicolumn{3}{c}{Pure Mergers Implementation} \\
    \munich  & $57^{+11}_{-8} \, \%$ & $14^{+52}_{-14} \, \%$ \\
    K10 (classical bulges) & $80^{+15}_{-36} \, \% $ & $55^{+21}_{-23} \, \% $ \\
    \hline
  \end{tabular}
  \caption{Predicted and observed fractions $f_{\rm bl}(M_\star)$ of
    bulgeless galaxies around MW-like haloes. Theoretical predictions
    refer to the median $f_{\rm bl}(M_\star)$ value, with the
    confidence levels defined on the fith and $95^{th}$ precentiles of
    the distribution, while confidence levels for the K10 data are
    based on the Wilson score interval approximation.}
  \label{tab:frac}
\end{center}
\end{table}

In order to compare the results shown in fig.~\ref{fig:mass} with the
K10 data, we have computed the galaxy stellar mass for all galaxies in
the K10 sample. In particular, we have used table B1 from
\citet{Zibetti09}, the $K$-band magnitudes given in K10, and $(B-V)_0$
colours obtained using the HyperLeda database \citep{Paturel03}. Given
the tight relation between the stellar mass of central galaxies and
their parent halo mass, we expect this approach to be consistent with
the K10 analysis (which is based on rotation velocities), and it
provides a more straightforward comparison with our theoretical
predictions. We thus compute the fractions of bulgeless galaxies
$f_{\rm bl}(M_\star)$ in two mass bins ($10^{10} < M_\star/M_\odot <
10^{11}$, and $10^{11} < M_\star/M_\odot < 10^{12}$), by using the
$B/T$ ratios listed in K10 (see their table~2). We give our estimated
fractions in Table~\ref{tab:frac}, together with the Wilson score
intervals\footnote{The \citet{Wilson27} approximation provides an
  estimate for asymmetric errors based on confidence levels of a
  binomial distribution, which is more appropriate than a normal
  distribution for small number statistics.} calibrated at the $95\%$
confidence limit.

In order to take into account the error due to cosmic variance, we
consider all central galaxies living in Milky Way-type haloes ($10^{12
  \pm 0.25} M_\odot$) in the \munich~model, and define K10-like
samples in the mass range $10^{10}<M_{\rm DM}/M_\odot<10^{12}$ by
considering all galaxies closer than 8 Mpc. We then compute the median
$f_{\rm bl}(M_\star)$ value and its confidence interval based on the
fifth and $95^{\rm th}$ percentiles of the distribution. This analysis
is limited to the \munich~model because \morgana~does not predict
accurate positions for satellite galaxies. In Table~\ref{tab:frac}, we
compare predictions for the standard \munich~run with the fraction of
disc galaxies in K10 sample (i.e. galaxies with total $B/T<0.1$), and
predictions obtained from the pure-merger model with the fraction of
galaxies with no classical bulge defined in K10 (i.e. classical
$B/T<0.1$). The theoretical fractions are systematically lower than
observations, and the discrepancy is severe for the standard
implementation compared to the distribution of pure discs: the
probability of observing a sample with the same morphological mix
measured by K10 is smaller than $1\%$ in the volume of the simulation
used by the \munich~model. The discrepancy between theoretical
predictions and observational estimates is, however, reduced if we
consider the predictions for the pure merger runs and compare them to
the fraction of galaxies without a classical bulge: in this case, the
probability of finding a MW-like neighbourhood similar to the K10
sample is larger than $5\%$. We note that when selecting MW-like
haloes, we have not applied any isolation criterion, which could
further increase the expected fraction of bulgeless systems. If
pseudo-bulges can be associated with secular processes, our results
clearly show the importance of the adopted modelling of disc
instabilities in order to correctly estimate the expected $f_{\rm
  bl}(M)$.
\begin{figure}
  \centerline{ \includegraphics[width=9cm]{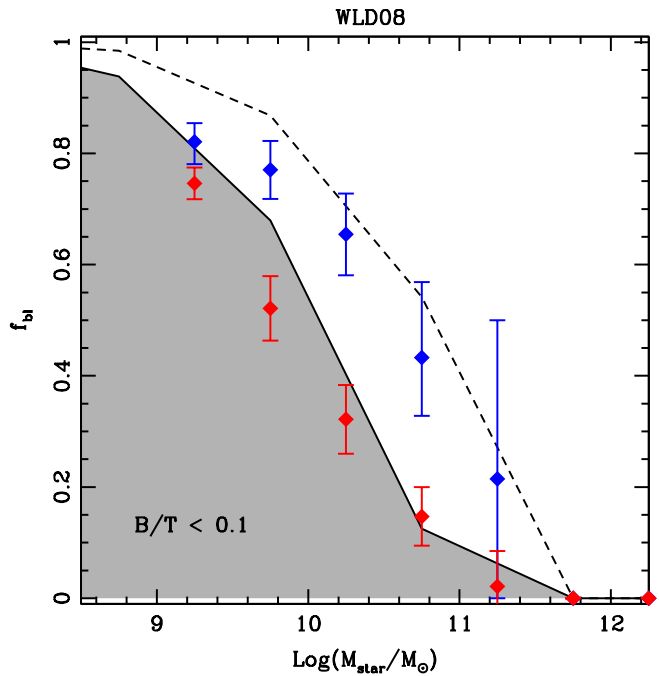} }
  \caption{The fraction $f_{\rm bl}$ of bulgeless galaxies as a
    function of stellar mass. Observational constrains from
    \citet{FisherDrory11} data are shown as solid (total $B/T<0.1$)
    and dashed lines (classical $B/T<0.1$) , while Red and blue dots
    refer to the standard and pure merger implementations of the
    \munich~model (see text for more details).}\label{fig:fd}
\end{figure}

We also compare predictions from theoretical models with the
observational estimates of \citet{Weinzirl09} and \citet{Cameron09}
who compiled samples that are larger than K10. However, the
\citet{Weinzirl09} sample is not mass and volume complete due to less
well controlled selection criteria, while the \citet{Cameron09}
decompositions suffer a number of additional problems, due to poorer
resolution of higher redshift objects: in fact we note that accurate
decomposition of galaxies is difficult, becoming worse at low
resolution (e.g. higher z) and where fewer components are included
(e.g. bars, see \citealt{Laurikainen07}). In addition, an accurate
comparison with these samples would require additional sources of
uncertainties in our model predictions, such as the inclusion of dust
attenuation and the modelling of synthetic SEDs.

Nonetheless, we make a qualitative comparison between data and models,
by defining a mass selected sample of $M_\star > 10^{10} M_\odot$
model galaxies, which roughly corresponds to the mass range covered by
observations. We find the following $z=0$ total stellar mass densities
for $B/T<0.1$ galaxies: $0.3$ and $0.9 \times 10^8 M_\odot Mpc^{-3}$
(for the standard and pure merger implementations of the
\munich~model, respectively) and $0.03$ and $0.6 \times 10^8 M_\odot
Mpc^{-3}$ (for the standard and pure merger implementations of
\morgana~model, respectively). These numbers are lower than those
found by \citet[$1.3 \pm 0.1 \times 10^8 M_\odot
  Mpc^{-3}$]{Cameron09}.

Using the same sample of model galaxies, we compare the predicted
fraction of $B/T<0.2$ high-mass spirals with the \citet{Weinzirl09}
results ($\sim66\%$, following that paper, we define spirals as
galaxies with $B/T<0.75$). These galaxies account for $18\%$ ($67\%$)
of objects in the standard (pure merger) implementation of the
\munich~model, and for $49\%$ ($81\%$) of objects in the standard
(pure merger) implementation of \morgana. These estimates confirm our
conclusion that bulgeless galaxies are underpredicted by the standard
implementations of semi-analytical models. The discrepancy is,
however, sensibly reduced when pure merger implementations of the same
models are considered.

In a recent publication, \citet{FisherDrory11} analysed the fraction
of classical and pseudo-bulges in the local volume (closer than 11
Mpc). We repeat our analysis on $f_{\rm bl}(M_\star)$, using this new
sample and the same approach as for the K10 sample
(fig.~\ref{fig:fd}). From the \citet{FisherDrory11} data (their
table.~1), we draw two subsamples including galaxies with both total
$B/T<0.1$ (solid line and gray shaded region) and classical $B/T<0.1$
(dashed line, we consider all galaxies showing a prominent
pseudo-bulge component as bulgeless), and we compare them with the
prediction of the standard (red diamonds with errorbars) and pure
merger (blue diamonds with errorbars) implementations of the
\munich~model, respectively. Samples of model galaxies are defined
within 11 Mpc from central galaxies of Milky Way-like haloes and we
compute the mean $f_{\rm bl}(M_\star)$ and its confidence interval
based on the $5^{\rm th}$ and $95^{\rm th}$ percentiles of the
distribution. In both cases, model predictions underpredict the
observational results; however, the discrepancy between the standard
implementation and the $f_{\rm bl}(M_\star)$ based on the total
$B/T<0.1$ ratio is reduced with respect to the K10 sample. The
agreement between the pure merger implementation and the sample based
on the classical $B/T<0.1$ is satisfactory. We note that theoretical
predictions obtained when considering the entire simulation box are
very close to the mean shown in fig.~\ref{fig:fd}. It is also worth
mentioning that the \citet{FisherDrory11} sample contains a
significant number of galaxies hosting a pseudo-bulge and with total
$B/T>0.3$ (they account for $\sim 10\%$ of the whole sample of
galaxies hosting a pseudo-bulge). Theoretically, we expect that such
prominent pseudo-bulges may hide a sub-dominant (hence relevant)
merger-driven classical bulge component and this effect would reduce
the discrepancy between the dashed line and blue diamonds.
\begin{figure}
  \centerline{ \includegraphics[width=9cm]{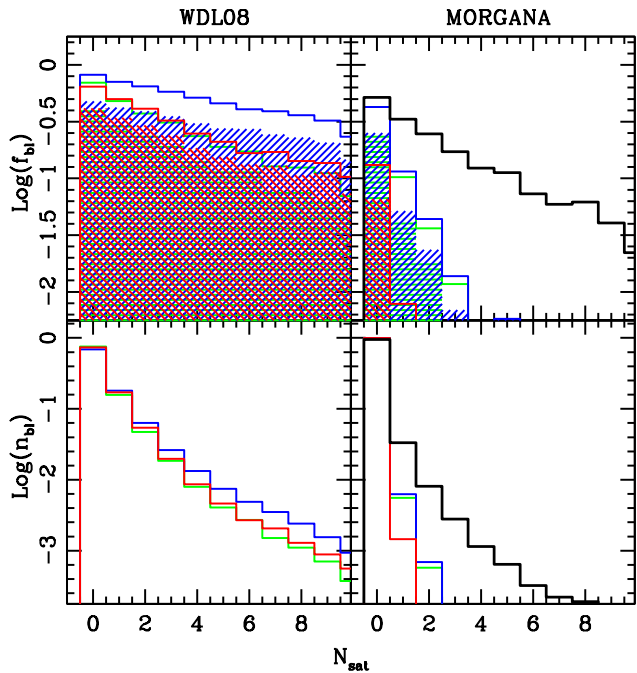} }
  \caption{{\it Upper panels}: $f_{\rm bl}$ as a function of the
    number of effective satellites (see text for more details). {\it
      Lower panels}: $n_{\rm bl}$ as a function of the number of
    effective satellites. In each panel colours, lines and shadings
    are the same as in fig.~\ref{fig:mass}.}\label{fig:sat}
\end{figure}

In order to provide limits on the number of mergers suffered by
bulgeless galaxies, we take advantage of the merger histories provided
by our models. In particular, for each galaxy in the model, we define
a number of ``effective satellites'' ($N_{\rm sat}$) as the number of
satellite galaxies in its parent halo, computed at the redshift when
it (its main progenitor) was for the last time a central galaxy. We
only consider satellite galaxies with $M_\star > 10^9 M_\odot$. For
central galaxies, the number of effective satellites corresponds to
the number of actual satellites. We show $f_{\rm bl}(N_{\rm sat})$ in
fig.~\ref{fig:sat} (upper panels). We note that the probability of
being bulgeless increases with decreasing $N_{\rm sat}$. The Figure
shows that \morgana~exhibits a very skewed distribution towards low
$N_{\rm sat}$ values, while the corresponding distribution for
\munich~is flatter. This is a consequence of the different merger
times adopted in the two models, as confirmed by the solid thick line
in the right panels that correspond to a run of the \morgana~model
(the pure merger run) with longer merger times. Since \morgana~assumes
(in its standard implementation) quite short merger times, only
galaxies living in very isolated environments survive as bulgeless
systems. The \munich~model adopts longer merger times, so that also
galaxies living in haloes with more substructures may avoid developing
a significant bulge. Disc instabilities do not change the
distributions, but as expected, they affect the fraction of bulgeless
galaxies. We stress that, despite a wider range of allowed halo merger
histories for bulgeless galaxies in \munich, the distribution of
$n_{\rm bl}(N_{\rm sat})$ (lower panels) is peaked towards very small
$N_{\rm sat}$ values.

\begin{figure}
  \centerline{ \includegraphics[width=9cm]{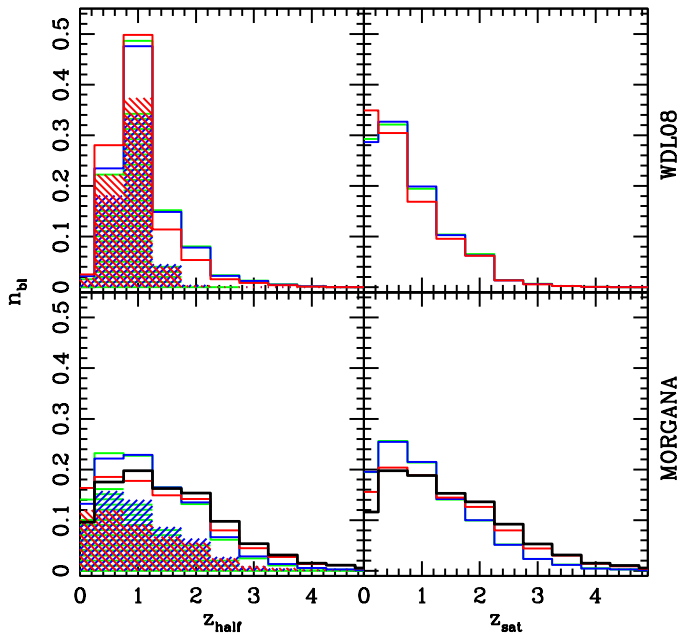} }
  \caption{{\it Left panel}: the distribution of formation redshifts
    of bulgeless galaxies (see text for more details).  {\it Right
      panel}: the distribution of accretion redshifts for bulgeless
    satellites. In each panel colours, lines and shadings are the same
    as in fig.~\ref{fig:mass}.}\label{fig:z}
\end{figure}

In order to better characterize the physical properties of bulgeless
galaxies as predicted from our models, we consider the quantity
$z_{\rm half}$, defined as the redshift at which half of the final
stellar mass is assembled in a single object. Since mergers play a
small role in the assembly of bulgeless galaxies, this quantity is
also a good indicator for the star formation history of the galaxy
(i.e. is a good approximation of the time when half of the stars were
{\it formed}). The distribution of $z_{\rm half}$ predicted by our
models is shown in fig.~\ref{fig:z} (left panels), and shows that
bulgeless model galaxies correspond to a relatively young population:
central galaxies are more skewed towards low $z_{\rm half}$ with
respect to satellites. These results are consistent with recent
observations of the central regions of M33, which appear to be
dominated by an old stellar population \citep[$\ge 6$ Gyr][]{Javadi11}
It is also worth noting that these objects contain a significant
fraction of old stars in their discs: the presence of 9-10 Gyr old
stars in a Milky Way-like galaxy is not unusual. Removing disc
instabilities or using the HOP09 merger prescription, does not modify
significantly the $z_{\rm half}$ distribution: this is due to the
paucity of bulge forming events in the bulgeless galaxies' history. We
also consider the distribution of accretion redshifts for the
bulgeless satellites ($z_{\rm sat}$, fig.~\ref{fig:z} right panels):
here we define $z_{\rm sat}$ as the last time the galaxy is the
central object of an independent DM halo. It is worth stressing that
this definition does not always correspond to the redshift when the
satellite is accreted onto the main progenitor of its $z=0$ parent
halo. Nevertheless, since $z_{\rm sat}$ the star formation history of
satellite galaxies is strongly affected by the strangulation in both
models. The Figure shows that the overall $z_{\rm sat}$ distribution
is broader than the $z_{\rm half}$ one. Therefore, both models
predicts a population of pure red discs in groups and clusters,
dominated by quite old stellar populations, which have been accreted
as satellites and have stopped forming stars relatively recently.

\section{Discussion \& Conclusions}
\label{sec:final}

In this paper we study the statistics, distributions and formation
histories of galaxies with bulge-to-total mass ratios $B/T<0.1$, as
predicted by theoretical models in the $\Lambda$CDM framework. This is
the second paper of a series: in the first paper \citep{DeLucia11}, we
have studied the formation of spheroids, and in a third paper we will
focus on a detailed comparison between model predictions and
observational data (Wilman et al, in preparation).

We consider two independently developed SAMs: the \citet{Wang08}
implementation of the ``Munich'' model, and the most recent
implementation of \morgana~\citep{Monaco07}. In addition, we consider
three different implementations of each model: a standard run, a pure
merger model, and a modified merger model based on the prescriptions
proposed by \citet{Hopkins09b} and based on recend hydrodynamical
simulations of galaxy mergers. The two models used in this study
include the same channels for bulge formation (i.e. galaxy-galaxy
mergers and disc instabilities).

Our results highlight that models predict a non-negligible fraction of
bulgeless $M_\star<10^{11}M_\odot$ galaxies at $z\sim0$. For all
models and implementations considered, the fraction of bulgeless
galaxies decreases rapidly with increasing stellar mass, and becomes
negligible for $M_\star>10^{12}M_\odot$. Bulgeless galaxies are either
central galaxies in low mass haloes, or satellites in groups and
clusters. They assemble their mass at relatively low redshifts, but
they can host quite old stellar populations. Given our assumption that
bulges form during mergers, these galaxies are bulgeless because they
had a relatively quiet merger history.  Therefore, bulgeless galaxies
are more likely to form in dark matter haloes hosting few satellites,
where galaxy merger rates are low enough to assure that the merger
channel is inefficient in forming a large bulge. The comparison
between results of the two models, highlights the importance of the
assumed prescription for merger times \citep[see also][]{DeLucia10}.

We compare our model predictions with observational results from the
\citet{Kormendy10} sample. The most interesting discrepancy between
models and data is found for galaxies with
$10^{10}<M_\star/M_\odot<10^{12}$, where the fraction of bulgeless
galaxies predicted by theoretical models is systematically lower than
the observational estimate. The discrepancy is significant when
comparing predictions from the standard model to the observed fraction
of `pure discs'. On the other hand, the predictions of the pure merger
implementations of our SAMs are statistically consistent (within
$95\%$ confidence level) with the observational estimates, if we
classify as bulgeless also galaxies with a relevant pseudo-bulge (as
done in the study by Kormendy et al.). Our results show that model
galaxies without a classical bulge are not rare: they can account for
up to $14\%$ of the total mass budget of galaxies in the range
$10^{11}<M_\star/M_\odot<10^{12}$. Our conclusions are confirmed by
the quantitative comparison with recent results by
\citet{FisherDrory11}, and by a qualitative comparisons of model
predictions with larger galaxy samples defined in \citet{Weinzirl09}
and \citet{Cameron09}. These results are in line with previous studies
in showing the importance of the adopted modelling of disc instability
to account for the estimated $B/T$ of our own Milky Way
\citep{DeLuciaHelmi08, Maccio10}.

We note that our conclusions depend on the assumed $B/T$ threshold for
the definition bulgeless galaxies. In particular, the fraction of
bulgeless galaxies decreases when adopting a lower threshold (and
viceversa). However, for $10^{10}<M_\star/M_\odot$ and
$M_\star/M_\odot>10^{12}$ our results do not significantly change when
varying the adopted threshold. This means that, for decreasing $B/T$
thresholds, the slope of the $f_{\rm bl}(M_\star)$ distribution
increases, and the $n_{\rm bl}(N_{\rm sat})$ distribution becomes more
skewed towards low $N_{\rm sat}$ values. Redshift distributions are
unaffected. The choice for the $B/T$ threshold may thus represent an
important aspect of our comparison between data and model
predictions. Given the observational uncertainties in the bulge/disc
decomposition \citep[e.g.][]{TascaWhite05, Laurikainen07}, we believe
that our $B/T<0.1$ choice is a reasonable one.

The main result of this paper is that SAMs predicts enough
merger-quiet galaxies to be (almost) consistent with the dearth of
classical bulges in the local Universe. In particular, we find that
the claimed tension between observational estimates based on the local
volume and predictions from SAMs weakens considerably when (i) disc
instability is switched off leaving mergers as the only channel to
create bulges in models, and (ii) only classical bulges are considered
in the data. Some caveats should be considered. The theoretical models
adopted in this paper are tuned to reproduce the observed fraction of
stellar mass locked in bulges at $z \sim 0$. If the instability
channel is switched off, these models might under-produce the stellar
mass in bulges with respect to the observational estimates. In
addition, as discussed extensively in Paper I, our modelling of disk
instability is very simplistic which has important consequences for
the formation and statistics of bulges\footnote{It is worth mentioning
  that the \citet{Bower06} model, which is not considered in this
  work, adopts yet different assumptions for instabilities: at each
  instability event, the whole disc is destroyed and all its baryons
  are given to the spheroidal component.}.

The models considered here do not predict the detailed properties of
discs (other than baryonic masses and scale-lengths), and they assume
no angular momentum dissipation, i.e. perfect conservation of the
angular momentum imprinted in the baryonic components (usually
modelled assuming some realistic distributions derived from numerical
simulations). Simulations show that this approach is probably
oversimplified, and that also haloes with a quiet merging history
acquire only a fraction of the angular momentum required for
rotational support by tidal torques (see e.g
\citealt{DOnghiaBurkert04}).

Finally, model bulges assemble their mass as the result of both
mergers and disc instabilities. Given the assumed tight connection of
classical (pseudo) bulges with mergers (disc instabilities), we expect
bulges to be composite systems. The K10 sample that we have used in
this study contains a relevant population (5 out of 19) of galaxies
hosting a relatively large ($0.1<B/T<0.4$) pseudo-bulge, and with no
evidence of a classical component. In order to use these observations
to constrain model predictions, it is of critical importance to
determine to which extent a massive pseudo-bulge may hide a classical
component and vice-versa. \citet{Gadotti09} results suggest that
``composite'' (actually classical bulges with signs of star formation
activity) bulges are indeed common in the local Universe:
unfortunately, the decomposition of a galaxy's photometry into the
contribution of different components is a challenging task, and an
unambiguous separation of classical and pseudo bulges is currently
possible only at low redshift (see also
\citealt{TascaWhite05}). Advances in this field will enable us to
increase our knowledge of the complex interplay of the physical
mechanisms responsible for the distribution of observed galaxy
morphologies.

\section*{Acknowledgements}
We are grateful to Jie Wang for letting us use the outputs of his
simulations. FF acknowledges the support of an INAF-OATs fellowship
granted on ``Basic Research'' funds. GDL acknowledges financial
support from the European Research Council under the European
Community's Seventh Framework Programme (FP7/2007- 2013)/ERC grant
agreement n. 202781. We acknowledge the usage of the HyperLeda
database (http://leda.univ-lyon1.fr)

\bibliographystyle{mn2e}
\bibliography{fontanot}

\label{lastpage}

\end{document}